Anomalous phonon behavior in the high temperature shape memory alloy: TiPd:Cr

S. M. Shapiro, G. Xu, B. L. Winn
Condensed Matter Physics/Materials Science Department
Brookhaven National Laboratory, Upton, NY 11973

D.L. Schlagel, T. A. Lograsso
Ames Laboratory and Materials Science and Engineering
Iowa State University, Ames, Iowa 50011

R. Erwin
NIST Center for Neutron Research
100 Bureau Drive, MS 8562
Gaithersburg, MD 20899-8562

ABSTRACT

$Ti_{50}Pd_{50-x}Cr_x$ is a high temperature shape memory alloy with a martensitic transformation temperature strongly dependent on the Cr composition. Prior to the transformation a premartensitic phase is present with an incommensurate modulated cubic lattice with wave vector of $q_0$=(0.22, 0.22, 0). The temperature dependence of the diffuse scattering in the cubic phase is measured as a function temperature for x=6.5, 8.5, and 10 at. %. The lattice dynamics has been studied and reveals anomalous temperature and q-dependence of the [110]-$TA_2$ transverse phonon branch. The phonon linewidth is broad over the entire Brillouin zone and increases with decreasing temperature, contrary to the behavior expected for anharmonicity. No anomaly is observed at $q_0$. The results are compared with first principles calculation of the phonon structure.

I. INTRODUCTION:

Shape memory alloys (SMA) are of considerable interest because of their technological applications ranging from actuators to medical stents[1]. The shape memory effect is also responsible for the superelastic properties of the materials, which are used in a variety of daily items from eyeglass frames to wire supports for brassieres. The fundamental character of SMA is that they exhibit a martensitic transformation where on cooling, the structure changes from a cubic structure to a lower symmetry structure with several coexisting variants or domains [2]. Most SMA have transformation temperatures near room temperature or below, which limits their applicability. There is a strong need for SMA with elevated transformation temperatures, which would open an entire area of applications in, for example, engine technology. The development of high temperature SMA has been slow because in many cases it is difficult to perform the standard studies such as transport properties at the elevated temperatures. Nevertheless several materials exist that undergo martensitic transformations at temperatures well above room temperature.

Ti-Pd is one such alloy that undergoes a martensitic transformation at about 800K for the equiatomic composition [3]. The high temperature phase is ordered and has the B2 CsCl-type crystal structure. Upon cooling it is reported that the crystal transforms displacively into a closed pack B19 (2H type) martensite [3]. By substituting a transition element such as V, Cr, Mn, Fe, or Co for Pd, the transformation temperature decreases [4,5] and an intermediate or premartensitic phase is present which can be viewed as a cubic phase with a superimposed modulation. The system chosen for study is TiPd-Cr because $T_M$ can be reduced to a more convenient working temperature [6,7]. Fig. 1 shows the phase diagram for Cr substitution, $TiPd_{1-x}Cr_x$ reproduced from the work of Enami et. al [4]. This diagram shows that for x=0 the transition is about 800K and decreases to about 350K for x=8. The high temperature austenite phase is the B2 structure and for x>5 at. % an incommensurate, intermediate, premartensitic phase is present where the lattice is modulated with a wavevector reported to vary with composition [6]. For x>4 at.% the martensite phase is a mixture of the B19 and 9R structures, while for x<4 at.% the martensite structure is B19. We have chosen to study the compositions x=6.5, 8.5 and 10 at. %. Preliminary reports of this work have already been published [8].

We report on elastic and inelastic neutron scattering experiments where the phonon dispersion curves are measured over a wide temperature range in the high temperature austenite B2 phase. The major result is that the [110]-$TA_2$ branch is heavily damped over the entire Brillouin zone. In contradiction to normal anharmonic behavior the linewidths decrease as the temperature increases indicating a mechanism, other than anharmoncity, is responsible for the damping. The most likely candidate is electron phonon interaction as evidenced by first principles calculation of the phonon behavior[9]

II. EXPERIMENTAL:

The crystals were grown at Ames Laboratory. The appropriate amounts of high purity Pd (99.989% by weight), Ti (99.99%) and Cr (99.995%) were arc melted several times under an inert atmosphere and then drop cast into a water chilled copper mold. A strain-anneal method was used to form the crystals, as the anneal method by itself did not produce sufficiently large crystals. The melting point was measured by DTA to be roughly ~1350°C, providing an upper limit for the heat treatment temperature. The ingot was then heat treated at 1275°C for 10 days for grain growth and coarsening. Grain boundaries were revealed by etching in $HNO_3$:HF 50:50 and the individual crystals were oriented by back reflection Laue and neutron diffraction. The chemical compositions were determined by scanning electron microscope energy dispersive analysis. Cr enters the Pd sites and the compositions of the alloys studied were $Ti_{49.7}Pd_{43.8}Cr_{6.5}$, $Ti_{50.5}Pd_{40.8}Cr_{8.5}$, $Ti_{49.6}Pd_{40.7}Cr_{9.7}$. The measured transition temperatures are in good agreement with those measured independently, which gives confidence to the composition values of Cr. The transition temperatures have also been plotted as a ratio of e/a in the literature [10]. The lattice parameter for x=6.5 at.% measured in the β-phase at 400K is 3.162Å. The single crystals were irregular shape with approximate volume of $1cm^3$. Figure 2 shows the rocking curve of the (1,0,0) Bragg peak measured below and above the martensitic transformation temperature. In the martensite phase there are several peaks equidistant from the main peak due to the domain configuration of the martensite phase. As the crystal is heated into the β-phase above $T_M$, the peaks coalesce into a single sharp peak, whose width determines the mosaicity of the sample. The mosaics of the crystals studied were all less than 1°.

The crystals were mounted with [HK0] in the scattering plane. This allowed us to measure the [110]-$TA_2$ transverse acoustic branch corresponding to atomic displacements along the [1$\bar{1}$0] direction. To measure other branches the 6.5 at. % crystal was mounted briefly in an [HKK] scattering plane. In a CsCl-type structure the stronger Bragg peaks are those with Miller indices such that (H+K+L) = 2n which would favor the (1,1,0) or (2,2,0) as the Brillouin zone (BZ) for measurement. However, Ti has a negative neutron scattering length and the most intense peaks satisfy the condition (H+K+L) = 2n+1. Therefore most of the measurements were performed in the (1,2,0) BZ.

The neutron experiments were performed on the BT9 triple axis instrument at the NIST research reactor. A fixed final energy of $E_F = 14.7$ meV was used throughout most of the experiment with PG(002) crystals serving as monochromator and analyzers. A number of scans were performed with $E_F = 30.5$ meV. A PG filter was placed before the detector to reduce the higher order contamination. The collimation chosen for most of the studies was 40-20-20-40, which gave a measured energy resolution of 0.65 meV, FWHM as determined by the incoherent scattering of the sample. Most experiments were performed in the neutron energy loss configuration.

The samples were placed in a high temperature closed cycle refrigerator or an A. S. Scientific Products vacuum furnace, which allowed measurements from 10 K up to 1273 K. Most of the studies were confined to the cubic B2-β phase.

III. RESULTS:

A. *Elastic Diffuse Scattering*

$Ti_{50}Pd_{50-x}Cr_x$ has been studied previously by electron diffraction and high resolution transmission electron microscopy [6,7]. These studies revealed that in the composition range of 6< x <14 at. % Cr some of the atoms are shifted from the cubic sites by a transverse lattice displacement wave resulting in an incommensurate structure with onset temperature shown in the phase diagram of Fig. 1. Elastic scans on the x=6.5 at. % sample were performed along the $[1\bar{1}0]$ direction measured from the (1,2,0) Brillouin zone center. An intensity map of these scans is shown in Fig. 3. A ridge of scattering is seen along the $[1\bar{1}0]$ direction with a bright spot at q~0.22 and intense scattering near the (1,2,0) BZ center. Figure 4 is a scan along this direction extending from the (1,2,0) BZ center to the (0,3,0) BZ center. Two satellite peaks are present with different intensities but at the same q=0.22 relative to the BZ centers. Measurements on the 10 at. % and the 8.5 at. % alloys show the same q=0.22 satellites as for the x=6.5 at. % alloy, implying that the modulation is composition independent, in variance with the earlier electron diffraction study [6]. Figure 5 compares the temperature dependence of the satellite peak intensities measured on the x=6.5 and 10 at. % samples on cooling. The intensity of both begins to increase near T~500 K. The x=6.5 at.% sample increases more rapidly, reaches a maximum at T=360K and then decreases, but is still present at the lowest temperature measured. Accompanying this decrease is a change of symmetry as evidenced by the splitting of the Bragg peaks of the cubic phase due to a reduction of symmetry [Fig. 2]. The x=10 at. % shows a continuous increase on cooling and no change of the Bragg peak down to 10K. Thus, aside from the modulation developing there is no martensitic transiton occurring in this alloy. Figure 6 shows the inverse correlation length determined from the linewidths of the modulation peak as a function of temperature for the x=6.5 and 10 at.% samples. The width is always larger than the resolution, as measured by a transverse scan through the (1, 2, 0) Bragg peak, which indicates that there is no true long range order. For x=6.5 at. % the linewidth decreases as the transition temperature approaches and reaches a minimum near Tm and then increases in the Martensitic phase with decreased intensity. For x = 10 at. % sample the linewidth decrease linearly with T at high temperatures and below 400K it is temperature independent.

As shown in Fig. 3 and Fig. 4 a considerable amount of diffuse scattering is also present at small q-values near the Bragg peaks. Imaging this scattering via TEM usually results in a temperature dependent tweed pattern considered a precursor to the martensitic transformation [11]. The temperature dependence of this intensity is shown in Fig. 7 measured at q=0.05[110] near the (1,2,0) BZ center for the 10 at. % alloy. On cooling from 1200 K, the intensity increases, reaches a maximum near 800K, and then continues to increase on lowering the temperature. This is distinctly different from the intensity behavior of the satellite shown in Fig. 5 for x=10 at. % where the intensity begins to appear around 500K and increases monotonically down to the lowest temperature. The surprising observation is that 800 K is the martensitic temperature of the 50:50 TiPd alloy as shown in the phase diagram shown in Fig. 1.

B. *Phonon Behavior*

Before we describe the search for soft mode behavior in the TiPd:Cr system we present the phonon dispersion curves measured along two symmetry directions. Figure 8 shows the low energy portion of the dispersion curves of the acoustic mode measured along the [100] and [110] directions for the 6.5 at. % Cr alloy. Most of the measurements were performed at T=450K in the cubic phase except for the [110]TA$_2$ branch which was measured at T=873K. This will be discussed in more detail later. The error bars indicate the accuracy of the measurements and their orientations indicate the direction of the scans, except for the TA$_2$ branch where the error bars indicate the line widths of the observed spectral peaks. From the slope of the dispersion curves near q=0 one can determine the velocity of sound, $v_i$, and therefore the elastic constants, $C_i$, from the relationship $\rho v_i^2 = C_i$ where $\rho$ is the density of the solid. In a cubic system there are 3 independent elastic constants determined from the slopes of the various branches. For this alloy the elastic constants are given in Table I. Also shown in the Table is A, the elastic anisotropy, which is the ratio of $C_{44}/C'$. This is determined from the slope of two transverse modes, TA$_1$ ($C_{44}$) and TA$_2$ (C') propagating along the [110] direction.

C. *Temperature and q-Dependence of [110]-TA$_2$ mode:*

Most of the experimental studies concentrated on measurements of the [110]-TA$_2$ branch since these modes exhibit anomalous behavior in nearly all bcc metals as predicted by Zener many years ago [12]. For q ~ 0 all the atoms are vibrating in phase and the phase of the different planes vary as q increases. In a number of previously studied systems such as Ni-Al [13], NiTi,[14] andNi$_2$MnGa [15], the phonon dispersion curves exhibit anomalous dips or kinks at the same wave-vector as the peaks in the elastic diffuse scattering. The dips are associated with a strong q-dependent electron-phonon coupling where the wave vector is determined by special spanning vectors of the Fermi surface and the electronic susceptibility is enhanced [16]. The elastic diffuse peaks are the static response of the charge density wave.

Measurements were made over a wide temperature range in both alloys to observe any dips in the dispersion curve of the [1$\bar{1}$0] - TA$_2$ branch at q=0.22, the wavevector of the peak in the diffuse scattering. The dispersion of the [110]-TA$_2$ in the austenite phase of the 6.5 and 10 at.% alloys are nearly identical and show very similar temperature dependence even though the 10 at.% alloy does not exhibit any phase transformation. Therefore, studies of the 10 at. % alloy permitted studies to much lower temperatures, unencumbered by a transformation to martensite. In both alloys the linewidth increases very rapidly with q and decreases with increasing temperature. This is shown in the spectra of Fig. 9. In Fig. 9a, measured at q =0.1, the phonon is quite broad at 500 K. On increasing the temperature the energy increases and the linewidth becomes narrower. On cooling further the linewidths increase and the phonon energies decrease such that the entire branch becomes ill defined. It is important to recall that no phase transformation occurs for this alloy and these observations are contrary to what is expected from a thermal induced anharmonic behavior where the opposite is expected. The q dependence of this branch also shows interesting behavior by comparing Fig. 9a and 9b. At 500 K, a peak is not present and the combined effect of a decrease in phonon energy and increase in linewidth gives an overdamped feature at q =0.2. On heating, a well-defined peak emerges. The temperature dependence of the energy and linewidth of two phonons measured at q=0.075 and 0.2 rlu is shown in Fig. 10a and 10b at elevated temperatures. For the smaller wavevector, the energy and the linewidth are nearly temperature independent. However for q=0.2, the energy decreases with decreasing temperature while the phonon linewidth increases

Another observation is that no phonon anomaly is associated with the q=0.22 wavevector where the diffuse scattering peaks exists. This can be seen in Fig. 11 where the phonon dispersion of the $TA_2$ branch measured at T=673K and T=1273K in the 10 at% alloy. The vertical bars and the shaded areas represent the measured linewidth (FWHM) for the two curves. Both curves show a linear increase at small q-vectors and then a turn over at higher q's. For T=673K, the curve starts to bend over around q=0.2, whereas at the higher temperature it bends over around q=0.3. The size of the damping for q>0.2 (T=673K) and q>0.3 (T=1273K) become comparable to the phonon energy, so the fits to the data are only approximate. For lower temperatures it is impossible to measure a dispersion curve since except for the very small q-values the spectra have an overdamped shape and the position is centered around E=0.

## IV. DISCUSSION:

As shown in the phase diagram of Fig. 1, the transitions temperatures determined in this experiment are consistent with those reported in the earlier study except that for the 10 at. % Cr, where no martensitic transition occurs. It is interesting to speculate whether this composition where the martensitic transition disappears could be the structural analogue of a quantum critical point observed in magnetic systems [17]. However all compositions studied show the appearance of a modulation of the B2 phase at elevated temperature with q-vector q[110], q=0.22, which is independent of composition. This differs from the TEM studies [6], which showed that q varies with composition from the commensurate value q=0.333 for 6 at. % Cr to 0.224 for 14 at. % Cr. This difference is not understood and suggests performing TEM studies on the same samples.

The inverse correlation length of the satellite peak is much broader than the resolution determined from a transverse scan through the Bragg peak and remains so over the entire temperature range studied. It is largest at the highest temperature and shows a decrease on cooling. For x=6.5 at. % the linewidth reaches a minimum near the transition temperature. For x=10 at. %, the linewidth decreases from high temperatures and then becomes constant below 400 K. This composition does not show a martensite transition. It is interesting to note that the break in the linewidth occurs near 400K, which is the very close to the transition temperature for x= 6.5 at. %. The correlation length varies from about 11 unit cells along the [110] direction (25Å) at elevated temperatures to about 20 unit cells (40Å) at lower temperatures. It is independent of Cr composition

In addition to the sharp diffuse scattering peak at q=0.22[110] there is considerable diffuse scattering near the Brillouin zone center, shown in Fig. 3, that is strongly temperature dependent as shown in Fig. 7. This diffuse scattering is present in many alloys that show martensitic transformations and, when imaged, shows a characteristic tweed pattern considered a precursor to the martensitic phase transformation [11]. The intensity is defect related, called Huang scattering, and is induced by atomic displacements surrounding defects in a solid. It is beyond the scope of this paper to study this extensively, but the intensity is usually anisotropic and inversely proportional to the elastic constants. In many bcc materials it is strongest along the [110]-$TA_2$ direction because along this direction, the elastic constant C', is anomalously low compared to the other elastic constants (see Table I) and the intensity is strongest. It is important to note that this scattering is totally elastic and due to static atomic displacements about a defect. The most surprising feature of Fig. 7 is the peak in this scattering near 800K even though this material does not show any transition. This temperature, 800K, is very close to the transition temperature of the $Ti_{50}Pd_{50}$ alloy. We can understand this if we consider that if the small amount of Cr added to this material substitutes randomly there is a high probability of finding several adjacent unit cells where Ti atoms will have only near neighbor Pd atoms. Therefore, locally it will behave as the 50-50 alloy which transforms at 800 K.

As in most bcc-based systems the [110]-$TA_2$ branch has the lowest energy and the largest temperature dependence as recognized by Zener [12]. The elastic constants given in table I are similar to those of another high temperature shape memory alloy, NbRu, also measured by neutron scattering [18]. The elastic anisotropy, A=3.6, is also similar to NbRu (A=4.8). There are no experimental measurements of the elastic constants for TiPd alloys, but there is a theoretical calculation of the elastic constants of the B2 phase based upon first principles total energy calculations [19]. These values are also given in Table I. The calculated values are for T=0 and the measured values are for elevated temperatures. The calculated $C_{12}$ is very close to the measured value, but the measured $C_{11}$ is significantly larger than the calculated value. The discrepancy is even larger for $C_{44}$. The calculated C' is negative which implies that the cubic phase is unstable at T=0 K, consistent with the observation that TiPd undergoes a martensitic transformation at elevated temperatures and the B2-phase is not the ground state of the system.

Many systems exhibiting a martensitic phase transformation, such as Ni-Al [13], Ni-Ti [20], $Ni_2MnGa$ [15], AuCd [21], FePt [22], exhibit pre-martensitic phases, which can be viewed as a cubic phase with a superimposed modulation having atomic displacements along the [1$\bar{1}$0] direction, transverse to [110]. The [110]-$TA_2$ phonon dispersion branch measured along this direction shows an anomaly at the same wavevector as the modulation. These displacements are the pathways by which the crystals transforms from the parent phase to the product, martensitic, phase. As the intensity of the diffuse scattering grows, the anomalies become more pronounced with a phonon softening at the wavevector. This is analogous to the observations in ferroelectric materials where the structural phase transition is adequately described by the soft mode theory developed nearly 50-years ago [23]. In this theory, the dynamical displacements of the soft mode are those needed to transform the high temperature structure into the lower symmetry ferroelectric phase. The frequency of the soft mode tends to zero and the crystal becomes unstable for those displacements and spontaneously distorts to the low temperature structure. In martensitic transformations, the transition is 1$^{st}$ order, so the modes do not go to zero at the transition temperature, but they do decrease down to the transition temperature.

The search for soft modes in TiPd:Cr revealed some surprising features. The [110]-TA$_2$ phonon branch exhibits anomalous momentum and temperature dependence in that the modes are very broad over nearly the entire Brillouin zone at room temperature. The linewidths of the phonons increase with decreasing temperature at the same time that the energies decrease. Anharmonicity would result in the opposite behavior so another type of coupling has to give this result. A clue to the naure of this coupling comes from first principles studies of the electronic and phonon structure of TiPd [9] . This type of calculation has proven to be very successful in explaining the anomalous phonon behavior in a number of other systems exhibiting martenstic transformations and shape memory behavior such as NiAl [16], NiTi [24] and Ni$_2$MnGa [25]. In theses systems they accurately calculate the wavevector of the instability as due to nesting Fermi wavevectors and strong electron-phonon coupling, the precise ingredients of a charge density wave (CDW). The recent calculations on TiPd [9] reveal an interesting behavior of the phonon dispersion curves of the cubic B2 phase. These calculations for T=0 show that many of the acoustic branches have a negative frequency, which implies that the B2 phase is unstable at T=0 and a phase transformation occurs at finite temperatures. It also implies that the B2 phase is dynamically stabilized by anharmonic phonons and that large fluctuations and local distortions are present in the B2 cubic phase. Of particular interest is the behavior of the [110]-TA$_2$ phonon branch, where the calculations show negative frequencies throughout the Brillouin zone. In contrast to other systems, such as NiAl, NiTi or Ni$_2$MnGa for which there is a well defined wavevector where the phonon dispersion curve becomes negative, there is no special wavevector for TiPd. The negative frequency for the entire branch is indicative that the branch is anomalous and consistent with the observation of broadening over the entire branch.

In summary we have measured the composition and temperature dependence of the elastic and inelastic neutron scattering of the high temperature shape memory alloy Ti$_{50}$Pd$_{50-x}$Cr$_x$. The diffuse scattering reveals an incommensurate transverse modulation of the cubic structure with wavevector $q_o=0.22[110]$ that is composition independent, along with temperature dependent Huang scattering near the Bragg peak. The [110]-TA$_2$ phonon branch is anomalously broad over the entire BZ as suggested by first –principles calculations of the phonon dispersion curves, but shows no singular anomaly at $q_o$. The linewidth decreases with increasing temperature, which cannot be explained by normal anharmonicity.

ACKNOWLEDGEMENTS

We thank Lee Tanner for many stimulating discussions. Work at Brookhaven and Ames Laboratory is supported by the Office of Science, U. S. Department of Energy under Contract Nos. DE-AC02-98CH10886, DE-AC02-07CH11358, respectively. We acknowledge the support of the National Institute of Standards and Technology, U.S. Department of Commerce, in providing the neutron research facilities used in this work.

# REFERENCES

1. *Science and Technology of Shape Memory Alloys: New Developments*, K. Otsuka and T. Takeshita, Guest editors, MRS Bulletin **27**, 91 (2002).
2. *Shape Memory Metals*, K. Otsuka, C. M. Wayman eds. (Cambirdge University Press, Cambridge, UK, 1998).
3. H. C. Donkersloot and J. H. N. VanVucht, J. of Less Common Met. **20**, 83 (1970).
4. K. Enami and Y. Nakagawa in *Proc. Int'l. Conf. on Martensitic Transformation (ICOMAT-92)*, ed. By C. M. Wayman and J. Perkins (Monterey Inst. For Advanced Studies, Monterey, CA, 1993) p. 521.
5. N. M. Mateeva, V. N. Kachin, and V. P. *Sivokha, Stabil'nye I Metastabil'nye Fazovye Ravnovesiya v Metallicheskikh Sistemakh* (Nauka Publ., Moskva), p. 25 (1985) (in Russian).
6. A. J. Schwartz, S. Paciornik, R. Kilaas, and L. E. Tanner, J. of Microscopy **180**, 51 (1995).
7. A. J. Schwartz and L. E. Tanner, Scripta Met. **32**, 675 (1995).
8. B. L. Winn, et al. Appl. Phys. A **74** (Suppl.), S1182 (2002); S. M. Shapiro et. al., J. Phys. IV France **112**, 1047 (2003).
9. X. Huang, K. M Rabe and G. J. Ackland, Phys. Rev. B **67**, 024101 (2003).
10. H. Hosoda, K. Enami, A. Kamio and K. Inoue, J. of Intel. Mat. Sys. And Stru. **7**, 312 (1996).
11. L. E. Tanner, Phil Mag **14**, 111 (1966); D. Schryvers and L. E. Tanner, Ultramicroscopy **37**, 241 (1990).
12. C. M. Zener, Phys. Rev. **71**, 846 (1947).
13. S. M. Shapiro, B. X. Yang, Y. Noda, L. E. Tanner and D. Schryvers, Phys. Rev. B **44**, 9301 (1991).
14. S. K. Satija, S.M. Shapiro, M. B. Salamon and C. M. Wayman, Phys. Rev. B **29**, 6031 (1984).
15. A. Zheludev, S. M. Shapiro, P. Wochner and L. E. Tanner, Phys. Rev. B **54**, 15045 (1996).

16. G. L. Zhao and B. N. Harmon, Phys. Rev. B **45**, 2818 (1992).
17. Z. Fisk, Nature **424**, 505 (2003).
18. S. M. Shapiro, G. Xu, G. Gu, J. Gardner and R. W. Fonda, Phys. Rev. B **73**, 214114 (2006).
19. G. Bihlmayer, R. Eibler and A. Neckel, Phys. Rev. B **50**, 13113 (1994).
20. C. M. Hwang, M. Meichle, M. B. Salamon and C. M. Wayman, Phil. Mag. A **47**, 9 (1983).
21. T. Ohba, S. M. Shapiro, S. Aoki and K. Otsuka, Jpn. J. Appl. Phys. **33**, L1631 (1994).
22. J. Kästner et al., Eur. Phys. Journal B **10**, 641 (1999).
23. W. Cochran, Adv. Phys. **9**, 387 (1960).
24. G. L. Zhao and B. N. Harmon, Phys. Rev. B **48**, 2031 (1993); X. Huang, C. Bungaro, V. Godlevsky and K. M. Rabe, Phys. Rev. B **65**, 014108 (2001)
25. Y. Lee, J. Rhee and B. N. Harmon, Phys. Rev. B **66**, 054424 (2002); C. Bungaro, K. M. Rabe and A. DelCorso, Phys. Rev. B **68** 134104 (2003).

# FIGURE CAPTIONS

Figure 1: Phase diagram of $Ti_{50}Pd_{50-x}Cr_x$ reproduced from Enami and Nakagawa (Ref.4)

Figure 2. Rocking curve of (1,0,0) Bragg peak above and below the martensitic transformation

Figure 3. Intensity contour measured around the (1,2,0) Bragg peak showing the satellite peak at q=(0.22,0.22,0) and diffuse scattering near the Bragg peak. The intensity scale bar on the right is a logarithmic.

Figure 4. Elastic diffuse scattering measured from the (0,3,0) (BZ) center to the (1,2,0) BZ center

Figure 5. Temperature dependence of the satellite intensity for $Ti_{50}Pd_{50-x}Cr_x$ for x=6.5 at. % (top) and x=10 at. % (bottom)

Figure 6. Temperature dependence of the inverse correlation length determined from the linewidth of the satellite peak for $Ti_{50}Pd_{50-x}Cr_x$ for x=10 at. % (top) and x=6.5 at. % (bottom)

Figure 7. Temperature dependence of the diffuse scattering measured near the (1,2,0) Bragg peak for $Ti_{50}Pd_{50-x}Cr_x$, x=10 at. %. Note that 800K is the martensitic transformation temperature of $Ti_{50}Pd_{50}$.

Figure 8. Partial phonon dispersion curves measured along the [00q] and [qq0] direction. The direction of the error bars indicate the direction of the scans. The bars for the [qq0]-$TA_2$ branch indicate the phonon linewidths measured at 873 K

Figure 9. x=10 alloy. Observed spectra at T=500 K and 1273 K for [qq0]-TA2 branch for (a) q =0.1 and (b) q=0.2. The curves at 1273 K are displaced in (a) and (b) by 500 cts. and 200cts, respectively

Figure 10. Temperature dependence of the phonon energy (top) and linewidth (bottom) for two phonons measured at q=0.075 and q=0.2

Figure 11. The phonon dispersion of the [qq0]-$TA_2$ measured at T=673K (top) and T=1273K (bottom). The bars and the shading indicate the linewidths determined from a Gaussian fit to the spectra

Table I: Elastic Constants (GPa)

|  | This Expt. | Theory (Ref. 22) |
|---|---|---|
| $C_{11}$ | 188 | 109.3 |
| $C_{44}$ | 38 | 0.6 |
| $C_{12}$ | 167 | 160.9 |
| $C'=1/2(C_1-C_{12})$ | 10.6 | -25.8 |
| $A=C_{44}/C'$ | 3.6 | |

Density: $\rho=7.64$ gms/cm$^3$

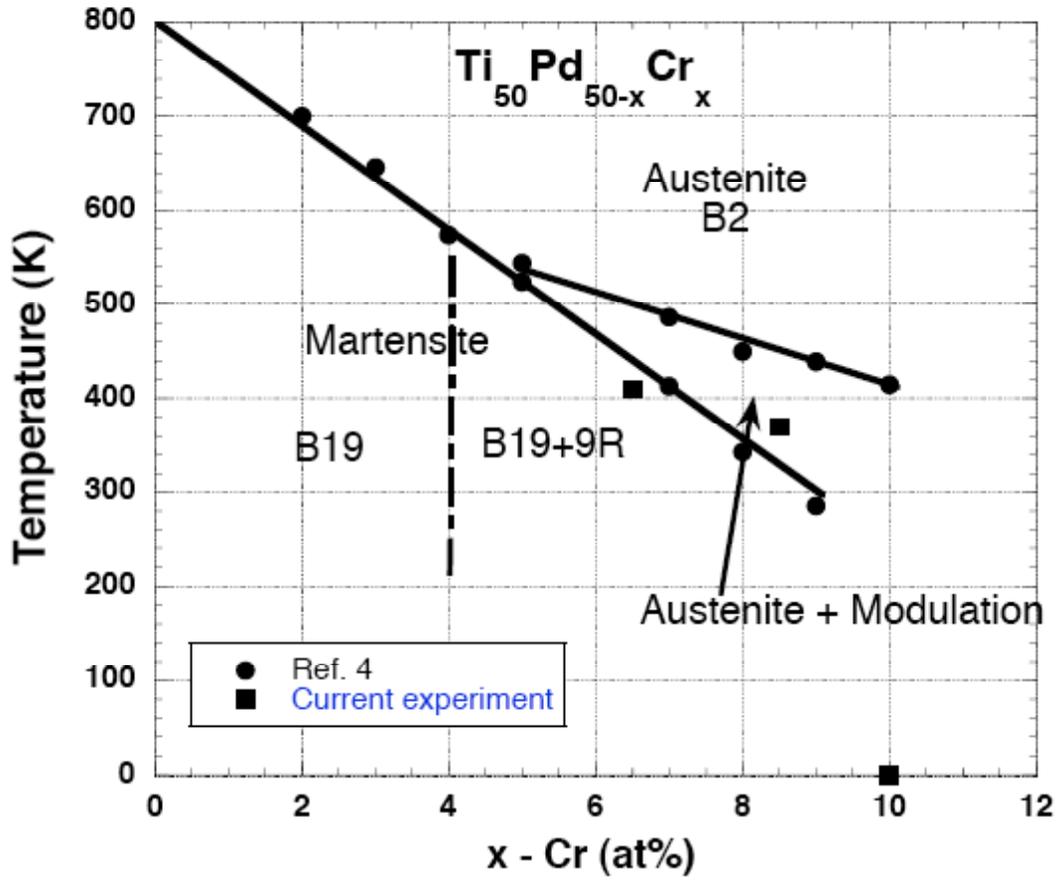

Figure 1 Phase diagram of $Ti_{50}Pd_{50-x}Cr_x$ reproduced from Enami and Nakagawa (Ref.4)

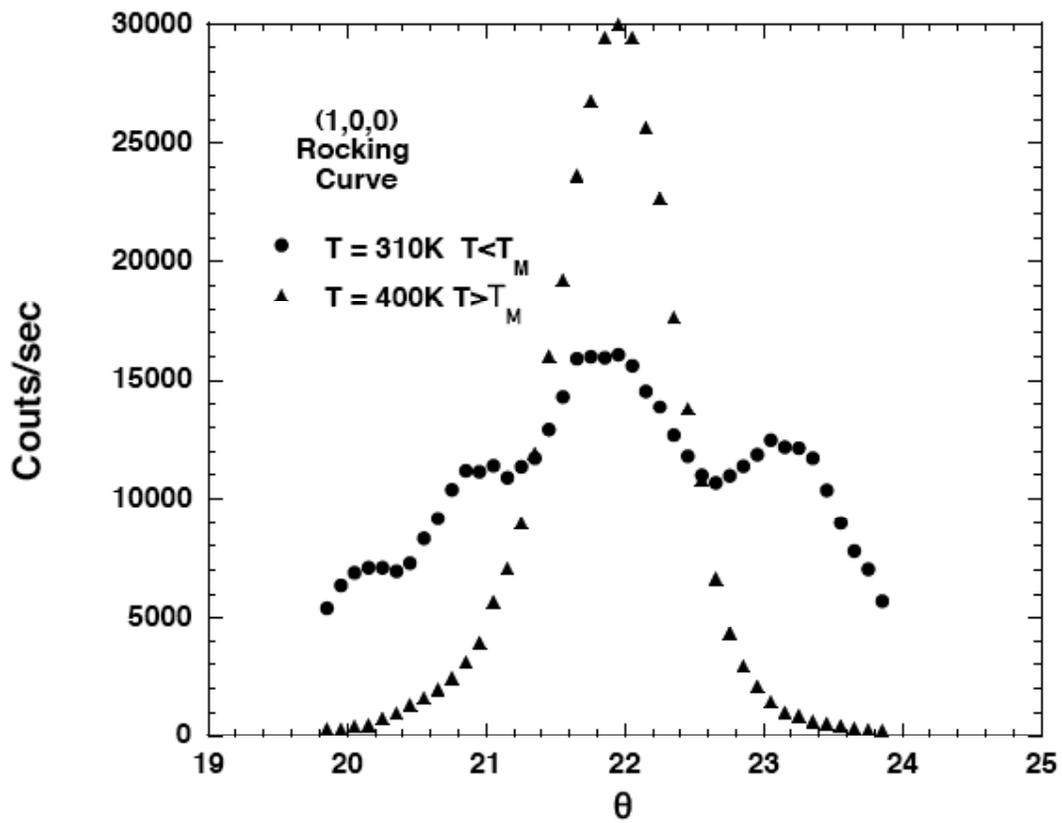

Figure 2. Rocking curve of (1,0,0) Bragg peak above and below the martensitic transformation for x= 6.5 at.% sample

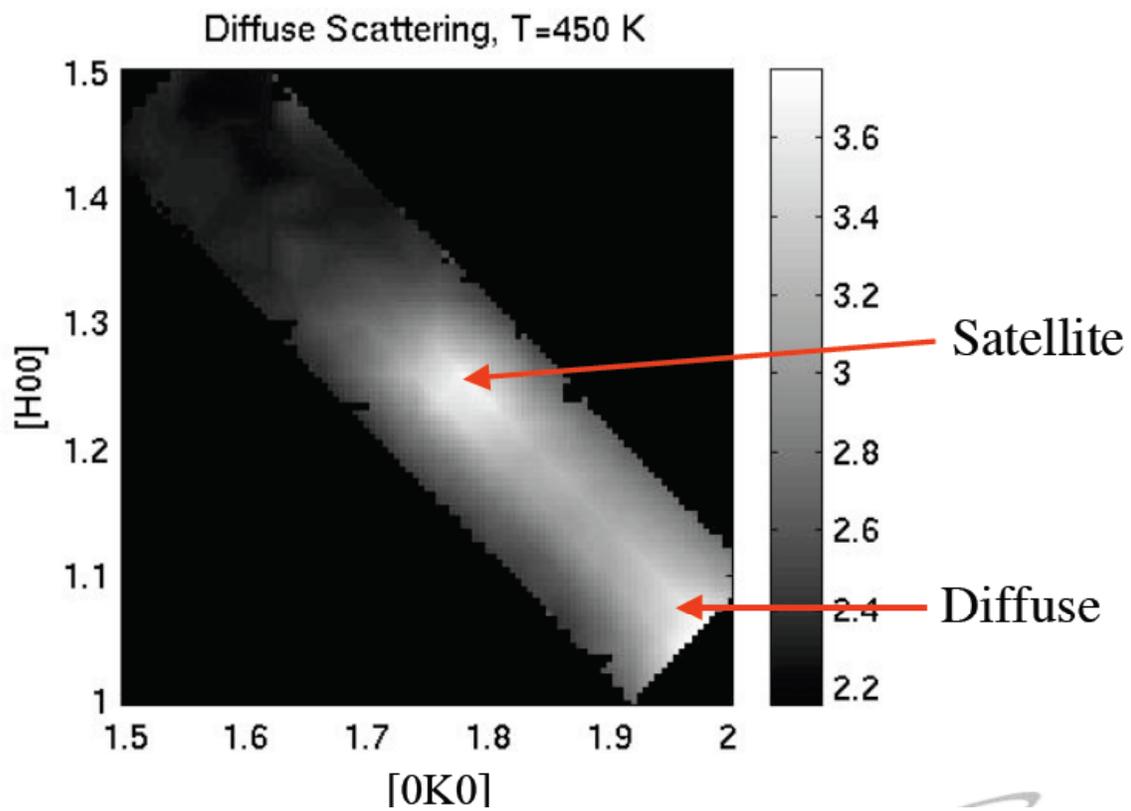

Figure 3. Intensity contour measured around the (1,2,0) Bragg peak for the 6.5 at. % alloy showing the satellite peak at q=(0.22,0.22,0) and diffuse scattering near the Bragg peak

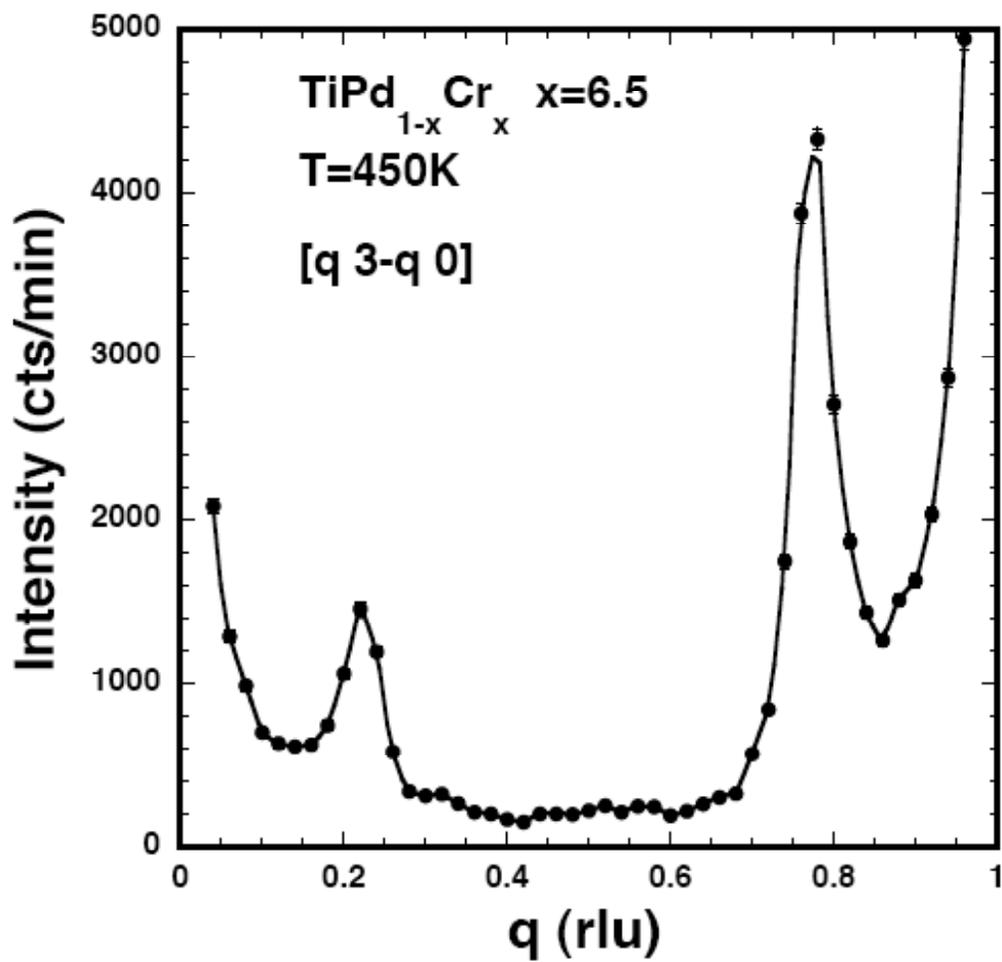

Figure 4. Elastic diffuse scattering measured from the (03,0) (BZ) center to the (1,2,0) BZ center

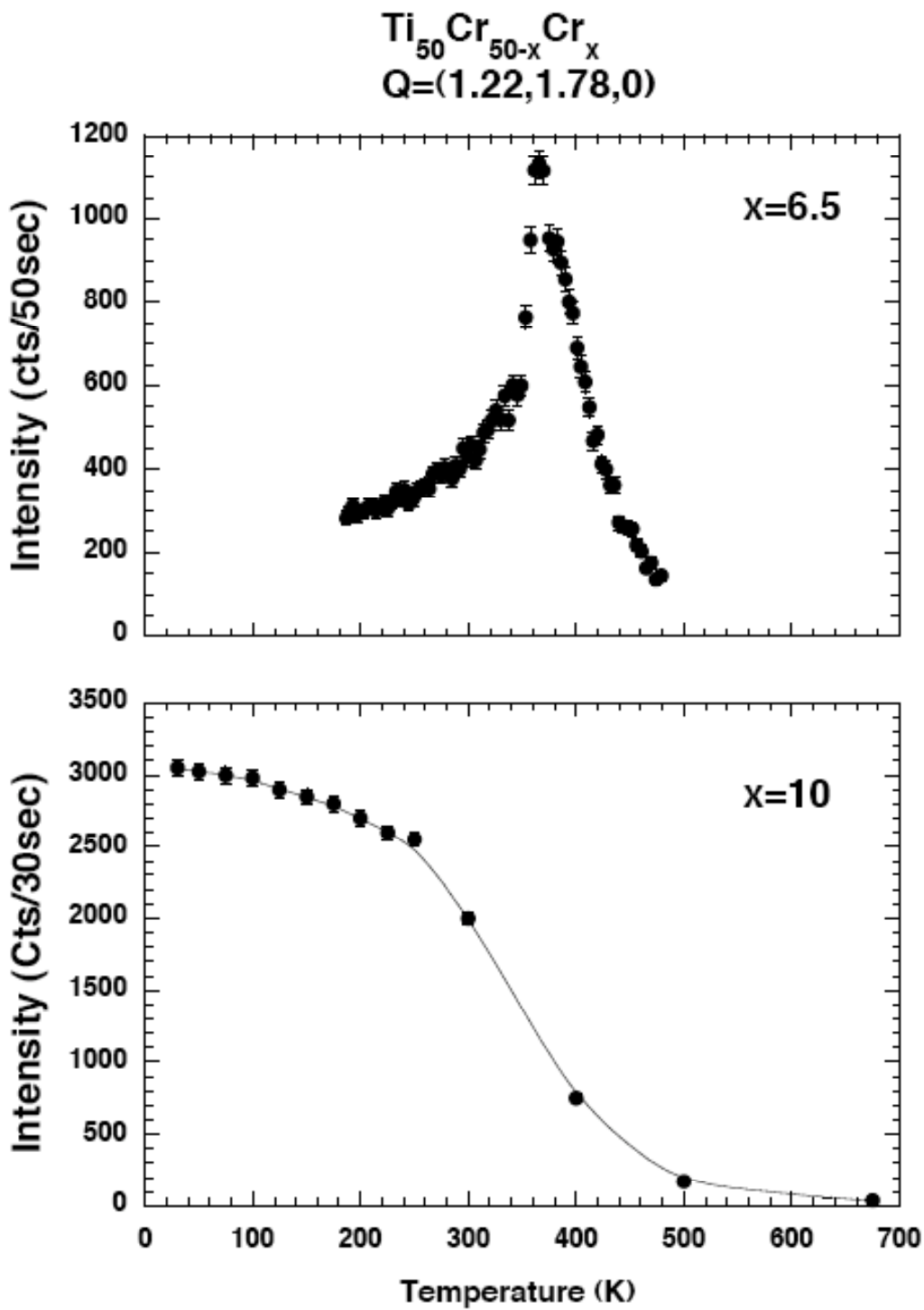

Figure 5. Temperature dependence of the satellite intensity for $Ti_{50}Pd_{50-x}Cr_x$ for x=6.5 at. % (top) and x=10 at. % (bottom)

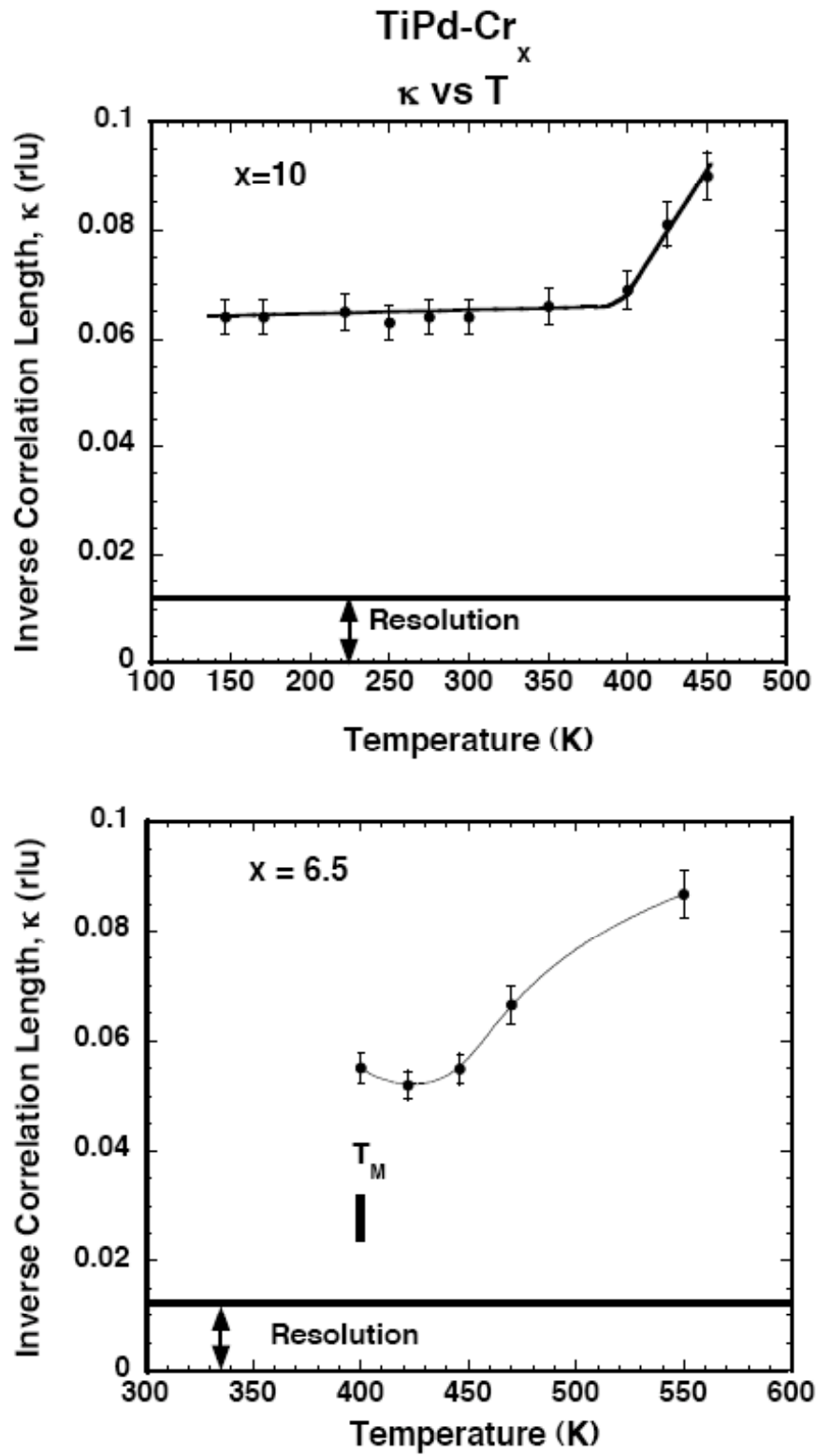

Figure 6. Temperature dependence of the inverse correlation length determined from the linewidth of the satellite peak for $Ti_{50}Pd_{50-x}Cr_x$ for x=10 at. % (top) and x=6.5 at. % (bottom)

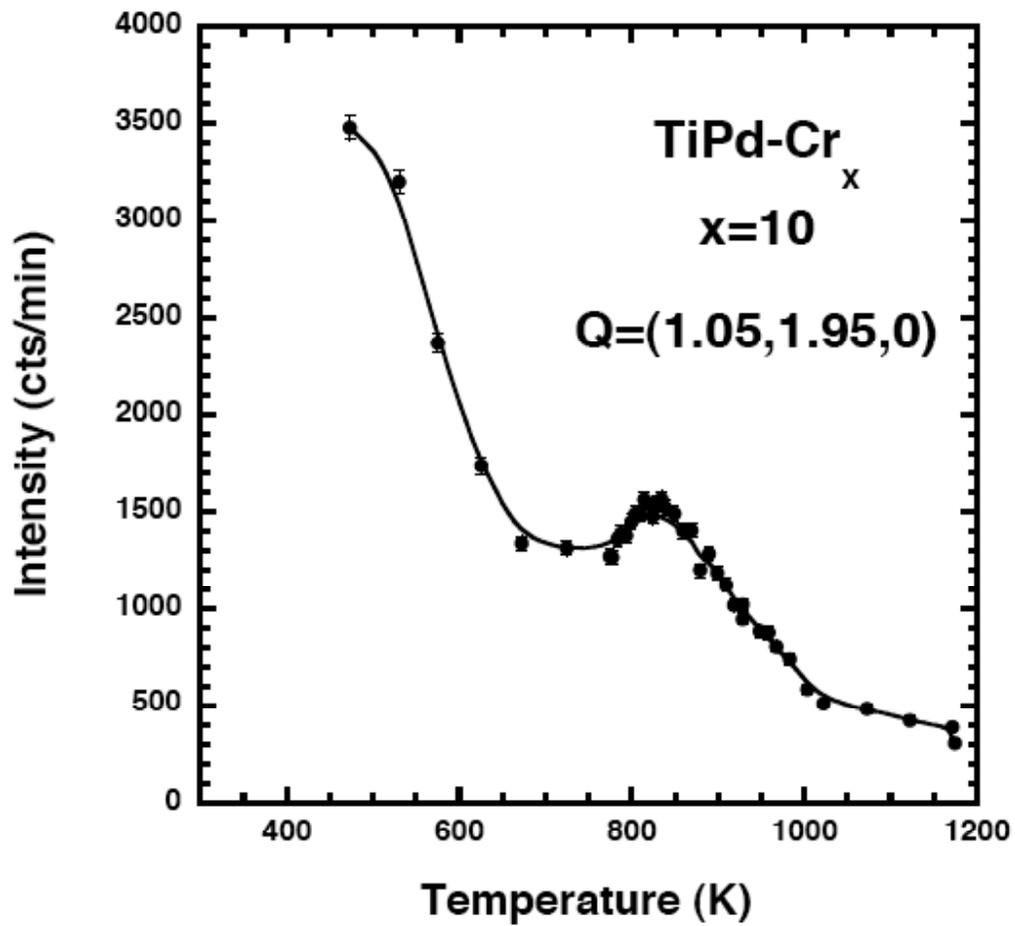

Figure 7. Temperature dependence of the diffuse scattering measured near the (1,2,0) Bragg peak for $Ti_{50}Pd_{50-x}Cr_x$, x=10 at. %. Note that 800K is the martensitic transformation temperature of $Ti_{50}Pd_{50}$.

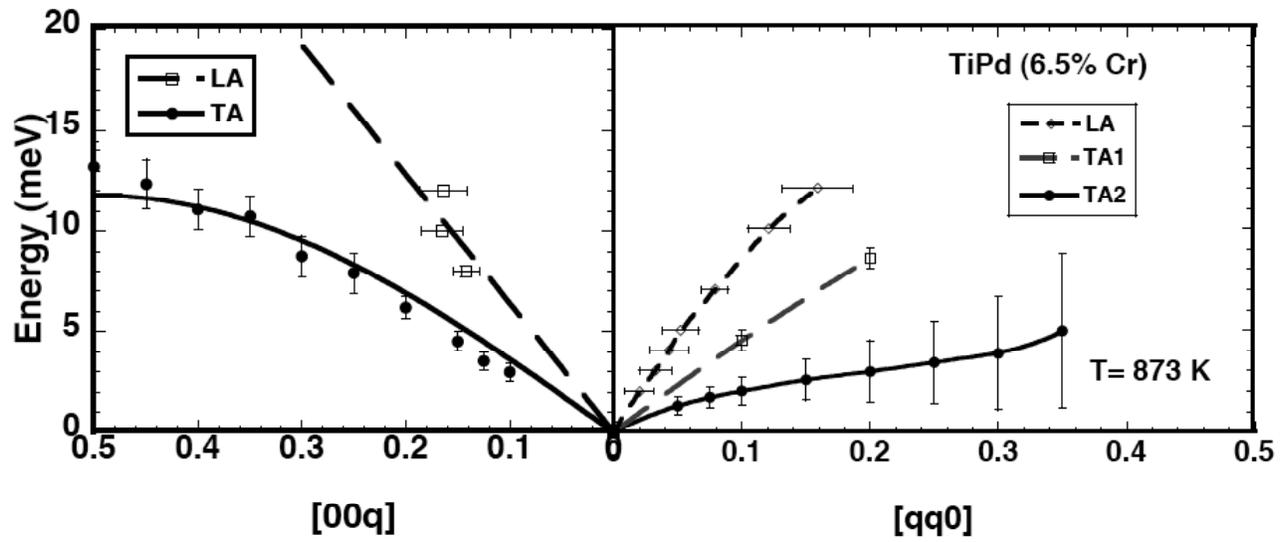

Figure 8. Partial phonon dispersion curves measured along the [00q] and [qq0] direction. The direction of the error bars indicate the direction of the scans. The bars for the [qq0]-$TA_2$ branch indicate the phonon linewidths measured at 873 K.

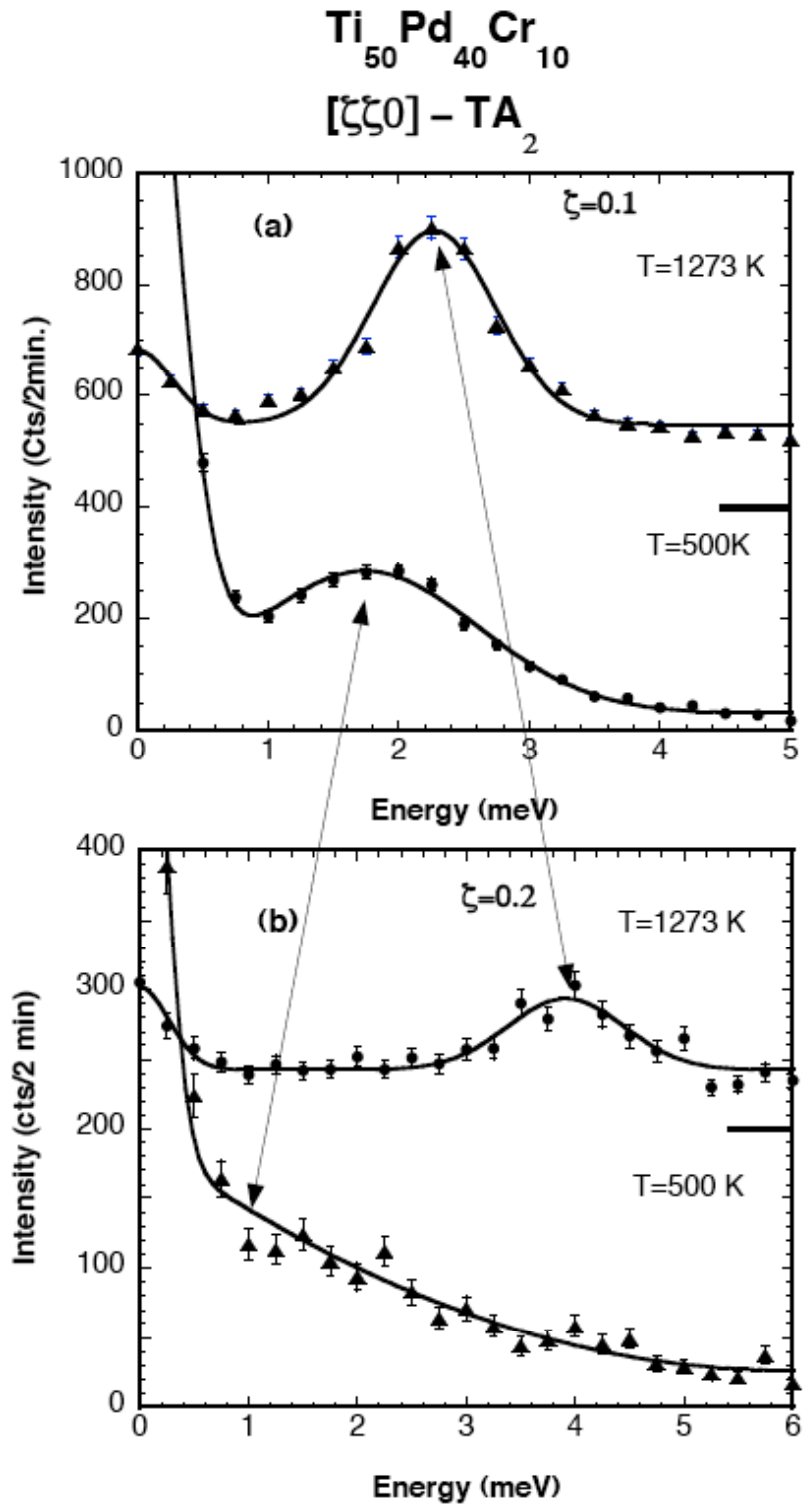

Figure 9. x=10 alloy. Observed spectra at T=500 K and 1273 K for [ζζ0]-TA2 branch for (a) ζ=0.1 and (b) ζ=0.2. The curves at 1273 K are displaced in (a) and (b) by 500 cts. and 200cts, respectively

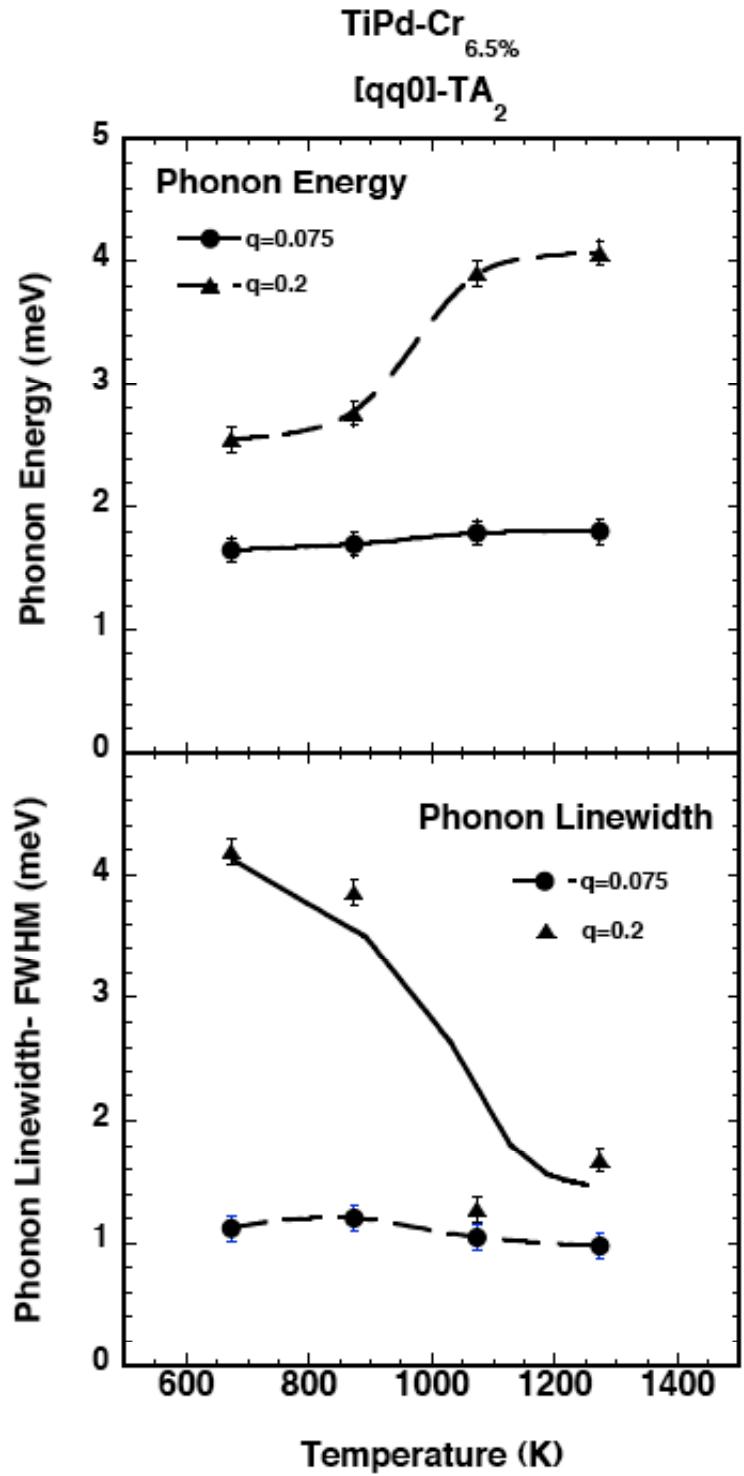

Figure 10. Temperature dependence of the phonon energy (top) and linewidth (bottom) for two phonons measured at q=0.075 and q=0.2

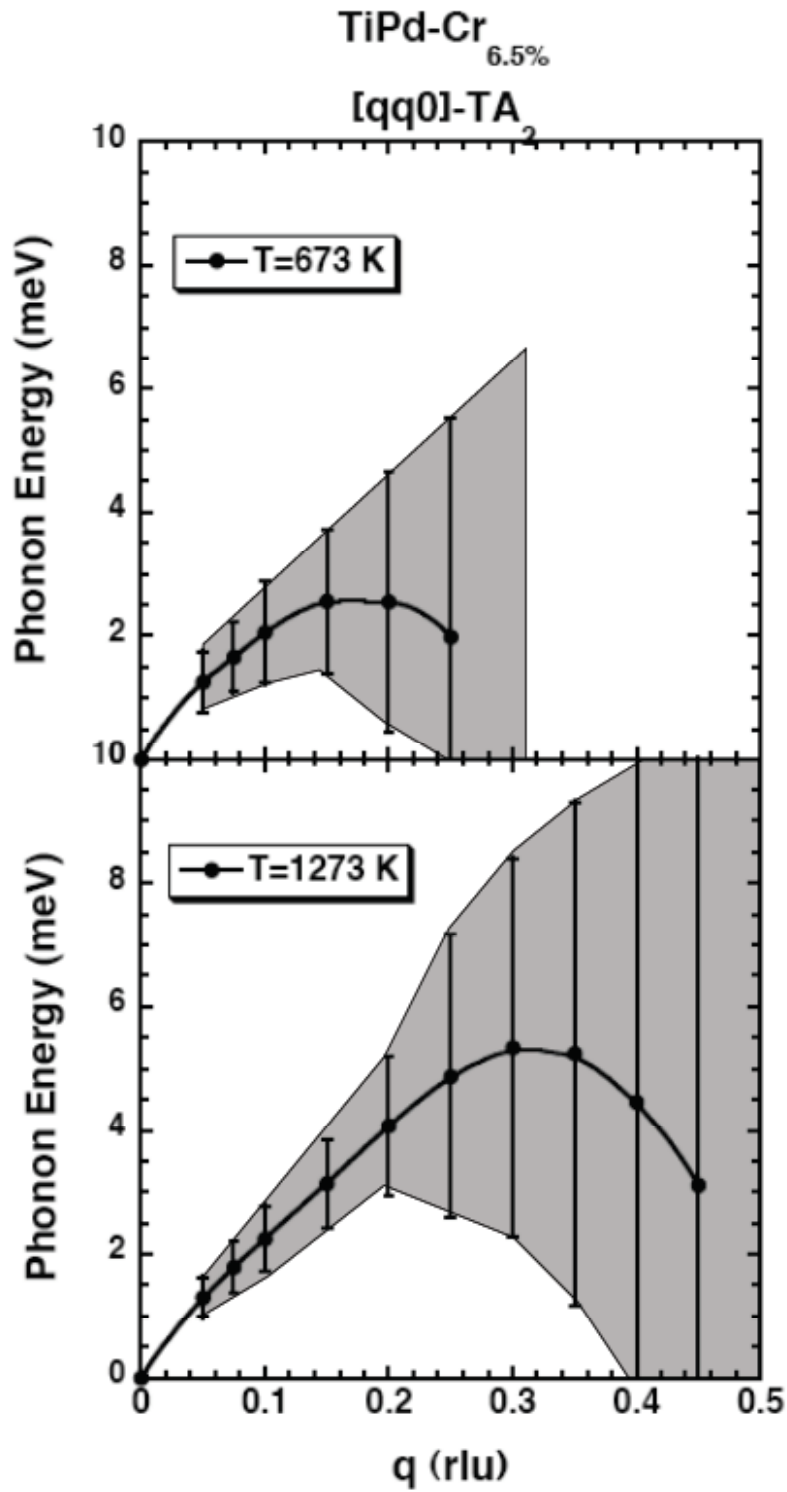

Figure 11. The phonon dispersion of the [qq0]-$TA_2$ measured at T=673K (top) and T=1273K (bottom). The bars and the shading indicate the linewidths determined from a Gaussian fit to the spectra

26